\documentclass[%
 reprint,
superscriptaddress,
 amsmath,amssymb,
 aps,
]{revtex4-1}
\usepackage{graphicx}
\usepackage{dcolumn}
\usepackage{bm}

\usepackage[utf8]{inputenc}
\usepackage[T1]{fontenc}
\usepackage{mathptmx}
\usepackage{etoolbox}

\makeatother
\begin{document}


\title{Temperature Sensing with RF-Dressed States of Nitrogen-Vacancy Centers in Diamond}
\author{Hibiki Tabuchi}
\email{hibiade0914@keio.jp}
\affiliation{Faculty of Science and Technology,
Keio University, Hiyoshi, Kohoku-ku, Yokohama 223-8522, Japan}
\affiliation{Center for Spintronics Research Network, Keio University, 3-14-1 Hiyoshi, Kohoku-ku, Yokohama 223-8522,Japan}
\author{Yuichiro Matsuzaki}%
\email{matsuzaki.yuichiro@aist.go.jp}
\affiliation{%
National Institute of Advanced Industrial Science and Technology (AIST), Umezono, Tsukuba, Ibaraki 305-8568, Japan 
}%
\author{Noboru Furuya}
\affiliation{%
Faculty of Science and Technology
Keio University, Hiyoshi, Kohoku-ku, Yokohama 223-8522, Japan
}%
\author{Yuta Nakano}
\affiliation{%
Graduate School of Natural Science and Technology, Kanazawa University, Kanazawa, Ishikawa 920-1192, Japan
}%
\author{Hideyuki Watanabe}
\affiliation{%
National Institute of Advanced Industrial Science and Technology (AIST), Umezono, Tsukuba, Ibaraki 305-8568, Japan 
}%
\author{Norio Tokuda}
\affiliation{%
Nanomaterials Research Institute, Kanazawa University.
}%
\affiliation{%
Graduate School of Natural Science and Technology, Kanazawa University, Kanazawa, Ishikawa 920-1192, Japan
}%
\author{Norikazu Mizuochi}
\affiliation{%
Institute for Chemical Research, Kyoto University, Gokasho, Uji-city, Kyoto 611-0011, Japan
}%
\author{Junko Ishi-Hayase}
\email{hayase@appi.keio.ac.jp}
\affiliation{%
Faculty of Science and Technology
Keio University, Hiyoshi, Kohoku-ku, Yokohama 223-8522, Japan
}%
\affiliation{%
Center for Spintronics Research Network, Keio University, 3-14-1 Hiyoshi, Kohoku-ku, Yokohama 223-8522, Japan
}%
\date{\today}

\begin{abstract}
Nitrogen vacancy (NV) centers in diamond are promising systems for realizing sensitive temperature sensors. Pulsed optically detected magnetic resonance (Pulsed-ODMR) is one of the ways to measure the temperature using NV centers. However, Pulsed-ODMR requires careful calibration and strict time synchronization to control the microwave pulse, which complicates its applicability. Nonetheless, the continuous-wave optically detected magnetic resonance (CW-ODMR) in NV centers is another more advantageous way to measure temperature with NV centers, owing to its simple implementation by applying a green laser and microwave in a continuous manner. This, however, has the drawback of a lower sensitivity compared to pulsed-ODMR. Therefore, to benefit from its accessible adaptation, it is highly important to improve the sensitivity of temperature sensing with CW-ODMR. Here, we propose a novel method to measure temperature using CW-ODMR with a quantum state dressed by radio-frequency (RF) fields under transverse magnetic fields. RF fields are expected to suppress inhomogeneous broadening owing to strain variations. Experimental results confirmed that the linewidth becomes narrower in our scheme compared to the conventional one. Moreover, we estimated the sensitivity to be approximately 65.5 $\mathrm{m}\mathrm{K}/\sqrt{\mathrm{Hz}}$, which constitutes approximately seven times improvement with respect to the sensitivity of the conventional scheme.
\end{abstract}

\maketitle

\section{Introduction}
As technology advances, the need to investigate matter at even smaller scales is continuously growing. It can be argued that thermal effects are more pronounced at very small and localized scales. Therefore, measuring the local temperature with high sensitivity and spatial resolution is of paramount importance to investigate properties of cells \cite{An2021,Kucsko,Yukawa2020} and nanodevices \cite{Andrich2018,Chen2021,Foy2020}. A nitrogen-vacancy (NV) center in diamond is a defect in which two adjacent carbon atoms are replaced by a nitrogen atom and a vacancy\cite{Doherty2013b,Rondin2014b}. Since its resonant frequency is dependent on temperature, an NV center is suggested as temperature sensor\cite{Acosta2010}. The NV center has a long coherence time of a few milliseconds \cite{Balasubramanian2009,Morishita2019,Herbschleb2019}, and could be incorporated in nanodiamond probes as small as tens of nanometers for high spatial resolution measurements. Moreover, as a temperature sensor, the NV center has a wide dynamic range from hundreds to thousands of Kelvin \cite{Fukami2019,Fujiwara2021,Liu2019}. Owing to these properties, the NV center is a promising candidate for realizing novel high-sensitivity and high-spatial resolution temperature sensors\cite{Andrich2018,Balasubramanian2009,Fujiwara2021,Fukami2019,Rondin2014b,Taylor2008}.

One of the ways to measure temperature with NV centers is pulsed-optically detected magnetic field (pulsed-ODMR) \cite{Choe2018,Neumann2013,Toyli2013,Wang2015}. A superposition of quantum states is created, where the change in temperature induced an additional phase in the superposition. The temperature can be estimated by detecting the phase shift via optical measurements. Although pulsed-ODMR generally provides high temperature sensitivity, it requires careful calibration, such as strict time synchronization to control the microwave pulse. 

The continuous-wave optically detected magnetic field (CW-ODMR) is another method for measuring temperature using NV centers. \cite{An2021,Kucsko,Tzeng2015,Yukawa2020}. In fact, temperature changes induces NV resonant frequency shifts, which can be measured from the spectrum. Temperature sensing by CW-ODMR is advantageous because it can be simply performed by the application of a green laser and microwave field in a continuous manner without complicated calibration. In addition, CW-ODMR is compatible with a charge-coupled device (CCD) that has a slow camera operation time and allows the collection of temperature information over a wide area from a single measurement \cite{Chen2021,Foy2020}. However, the sensitivity of CW-ODMR is typically lower than that of pulsed-ODMR. Therefore, it is important to improve the sensitivity of CW-ODMR for various practical applications.

In this study, we propose a novel method for measuring temperature using CW-ODMR with a quantum state dressed by radio-frequency (RF) fields under transverse magnetic fields\cite{Saijo2018,Yamaguchi2019}. RF fields are expected to suppress electric field noise. We performed CW-ODMR with RF fields under transverse magnetic fields and experimentally showed that the linewidth becomes narrower as we increase the intensity of the RF fields. From experimental results, we estimated the sensitivity of our scheme to be approximately 65.5 mK $/\sqrt{\mathrm{Hz}}$, which is seven times better than that of the conventional scheme.

\section{Principles and Experimental Results}
\begin{figure}[htbp]
  \centering
  \hspace*{-0.25cm}
  \includegraphics[width=9cm]{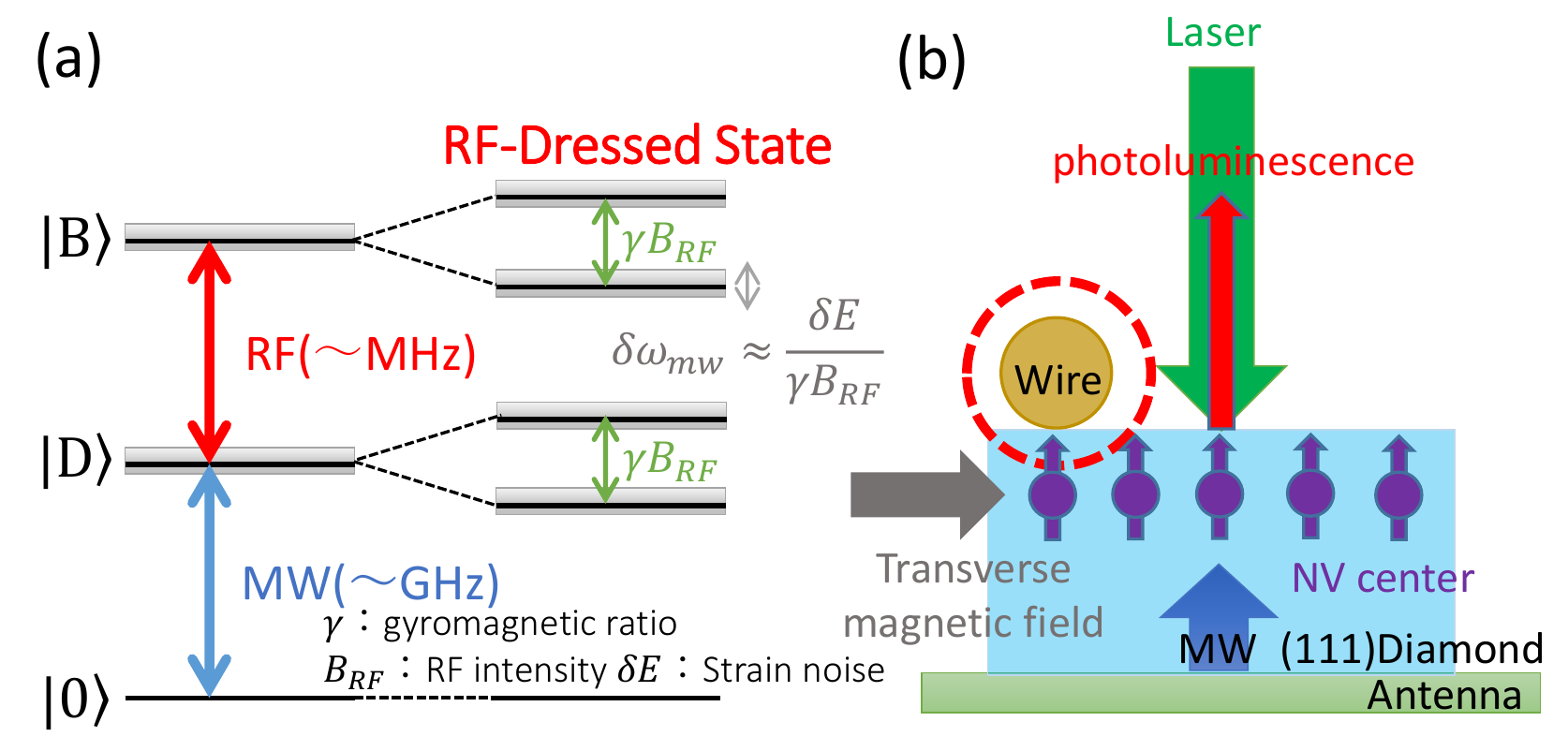}
  \caption{(a)The energy diagram of our system. We drive the NV centers by the radio-frequency (RF) and the microwave (MW) fields to generate and detect the dressed states. Here, $E_x$ denotes the strain or electric field, $\delta E$ denotes the strain (or the electric field) variations, $\delta \omega_{\mathrm{MW}}$ denotes the fluctuation of the resonance frequency fluctuation, $\gamma$ denotes the gyromagnetic ratio, $B_{\mathrm{RF}}$ denotes RF intensity, respectively. (b)Experimental setup to measure CW-ODMR.}
  \label{RFdressedstate}
\end{figure}

We shall describe in this section the physical properties of NV centers. As defined earlier, a nitrogen-vacancy (NV) center in diamond is a defect in which two adjacent carbon atoms are replaced by a nitrogen atom and a vacancy\cite{Doherty2013b,Rondin2014b}. The ground state of an NV center is a spin triplet and the energy eigenstates are spanned by $|0\rangle$ and $|\pm 1\rangle$\cite{Doherty2013b,Rondin2014b}. The NV center has four possible crystallographic axes, which align with a nitrogen atom and a vacancy. When magnetic fields parallel to the NV axis are applied, the degenerate level $|\pm 1\rangle$ is split into $|+ 1\rangle$ and  $|- 1\rangle$. On the other hand, when we apply strain or transverse magnetic fields, the degenerate level is split into  $|D\rangle=\frac{1}{\sqrt{2}}(|1\rangle-|-1\rangle)$ and $|B\rangle=\frac{1}{\sqrt{2}}(|1\rangle+|-1\rangle)$. When we apply a green laser to the NV centers, we initialize them in the $|\ 0\rangle$ state. Moreover, the spin state can be determined from the difference in the number of photons emitted from the NV centers. In addition, by sweeping the microwave frequency while applying a green laser, we observe a change in the emitted photon and can find the resonant frequency between $|\ 0\rangle$ and the other energy eigenstates. By applying a resonant microwave pulse, the spin state of the NV center can be controlled \cite{Dreau2011,Jensen2013}. Under the effect of strain or transverse magnetic fields, we can induce a transition between $|D\rangle$ and $|B\rangle$ by applying a resonant RF field \cite{Saijo2018,Yamaguchi2019}. In this case, the states $|D\rangle$ and $|B\rangle$ are split into other states dressed by RF fields, as shown in Fig. \ref{RFdressedstate}.

The Hamiltonian of the NV center with microwave and RF fields is as follows\cite{Saijo2018,Yamaguchi2019}: 
\begin{eqnarray}
 H_{\mathrm{NV}}=&D& \hat{S}_{z}^{2}+E_{x}\left(\hat{S}_{x}^{2}-\hat{S}_{y}^{2}\right)+E_{y}\left(\hat{S}_{x} \hat{S}_{y}+\hat{S}_{y} \hat{S}_{x}\right)+g \mu_{b} B_{x}\hat{S}_{x}\nonumber \\
 &+&\sum_{j=x, y, z} \gamma_{e} B_{\mathrm{MW}}^{(j)} \hat{S}_{j} \cos \left(\omega_{\mathrm{MW}} t\right)\nonumber\\
 &+&\sum_{j=x, y, z} \gamma_{e} B_{\mathrm{RF}}^{(j)} \hat{S}_{j} \cos \left(\omega_{\mathrm{RF}} t\right),
\end{eqnarray}
where $\hat{S}$ is the spin-1 operator of the electronic spin, $D$ is the zero-field splitting, $E_x$ ($E_y$) is the  strain along the x (y) direction, $g \mu_{b} B_{x}$ is the Zeeman splitting, $\gamma_{e}$ is the gyromagnetic ratio of the electron spin, and $B_{\mathrm{MW}}$ ($B_{\mathrm{RF}}$) is the microwave (RF) field intensity. Under conditions $D \gg g \mu_{b} B_{x} \gg E_{y}$, by going to a rotating frame with the rotating wave approximation, we obtain:
\begin{equation}
\label{E1}
H^{\prime}=\omega_{b} \hat{b}^{\dagger} \hat{b}+\omega_{d} \hat{d}^{\dagger} \hat{d}+J\left(\hat{b}^{\dagger} \hat{d}+\hat{b} \hat{d}^{\dagger}\right)+\lambda_{b}\left(\hat{b}+\hat{b}^{\dagger}\right).
\end{equation}
In Eq. \eqref{E1}, we have treated the system as a harmonic oscillator where $\hat{b}^{\dagger}$ ($\hat{d}^{\dagger}$) denotes a creation of the state of $|B\rangle$ ($|D\rangle$) because the ensemble of the NV centers was considered\cite{Yamaguchi2019}. 
We can also calculate the probability of the state being $|\ 0\rangle$ as follows:
\begin{eqnarray}
\label{E2}
p_0=&1&-|\frac{-\lambda_{b}\left(\omega_{d}-i \Gamma_{d}\right)}{\left(\omega_{b}-i \Gamma_{b}\right)\left(\omega_{d}-i \Gamma_{d}\right)-J^{2}}|^2\nonumber\\
&-&|\frac{\lambda_{b} J}{\left(\omega_{b}-i \Gamma_{b}\right)\left(\omega_{d}-i \Gamma_{d}\right)-J^{2}}|^2,
\end{eqnarray}
where $\omega_{b}=D+Ex-\omega_{\mathrm{MW}}$, $\omega_{d}=D+Ex-\omega_{\mathrm{MW}}+\omega_{\mathrm{RF}}$,$J=\frac{1}{2} \gamma_{e} B_{\mathrm{RF}}^{(z)}$, and $\lambda_{b}=\frac{1}{2} \gamma_{e} B_{\mathrm{MW}}^{(x)}$.
A sharp dip structure is observed in CW-ODMR with transverse magnetic fields or in absence of magnetic fields \cite{Dmitriev2019,hayashi2018optimization,Matsuzaki2016,Saijo2018,Yamaguchi2019,Zhu2014}, and this analytical solution can be used to fit the shape of CW-ODMR.

In the following, we shall explain the reasons for the robustness of the RF-dressed states to strain variations. From the above expressions, we can calculate the resonant frequency of the microwave as:
\begin{equation}
    \omega_{\mathrm{MW}}=\frac{1}{2}\left\{2 D \pm \omega_{\mathrm{RF}} \pm \sqrt{\left(2 E_{x}-\omega_{\mathrm{RF}}\right)^{2}+\left(\gamma_{e} B_{\mathrm{RF}}^{(z)}\right)^{2}}\right\}.
\end{equation}
By setting $\omega_{\mathrm{RF}}=2E_x$, we can generate RF-dressed states \cite{Saijo2018,Yamaguchi2019}.
Suppose that there are strain variations and the inhomogeneous width is described by $\delta E_{x}$. In this case, the resonance frequency fluctuates, and we obtain:
\begin{equation}
    \omega_{\mathrm{MW}}+\delta \omega_{\mathrm{MW}}=\frac{1}{2}\left[2 D \pm \omega_{\mathrm{RF}} \pm \sqrt{\left(\delta E_{x}\right)^{2}+\left(\gamma_{e} B_{\mathrm{RF}}^{(z)}\right)^{2}}\right],
\end{equation}
where $\delta \omega_{\mathrm{MW}}$ is the width of the fluctuation. When the RF intensity is sufficiently larger than the fluctuation of the strain, we can use the Taylor expansion to obtain:
\begin{equation}
\delta \omega_{\mathrm{MW}}\simeq \frac{1}{4} \frac{\left(\delta E_{x}\right)^{2}}{\gamma_{e} B_{\mathrm{RF}}^{(z)}}.
\end{equation}
This shows that by increasing $B_{\mathrm{RF}}^{(z)}$, we can suppress the effect of strain variations, and the linewidth in CW-ODMR becomes narrower.

Here, we describe our scheme for measuring the temperature using CW-ODMR with RF-dressed states. By performing CW-ODMR under the effect of transverse magnetic fields, the RF-dressed states become observable. We propose using the resonant frequency of these states to measure temperature. Importantly, as the temperature increases, the resonant frequency shifts owing to the change in the zero-field splitting $-74.2 \mathrm{~K} / \sqrt{\mathrm{Hz}}$; therefore, we should be able to detect such a shift using CW-ODMR. In our experiments, the RF-dressed states were robust against electric field noise. We confirm that as we increase the intensity of RF fields, the linewidth of the peak corresponding to the RF-dressed states becomes narrower. From experimental results of the CW-ODMR, we calculated the sensitivity of the temperature sensor.

For the experiments, we used a home-built system for confocal laser scanning microscopy and operated the electron spin of NV centers using a green laser (532 nm) and a microwave. The photoluminescence from the NV centers passes through a pinhole and is detected by an avalanche photodiode (APD). The diamond sample was positioned above the antenna \cite{Sasaki2016} used to emit microwaves. We applied RF fields by placing a copper wire on the diamond sample and adding an AC voltage. The NV axis in the diamond is aligned in a specific direction. The sample was grown by chemical vapor deposition (CVD) using N-doped diamond.  The density of the NV centers was approximately $10^{16}$ cm ${}^{-1}$.

\section{Result}
\begin{figure}[htbp]
  \centering
  \hspace*{-0.25cm}
  \includegraphics[width=9cm]{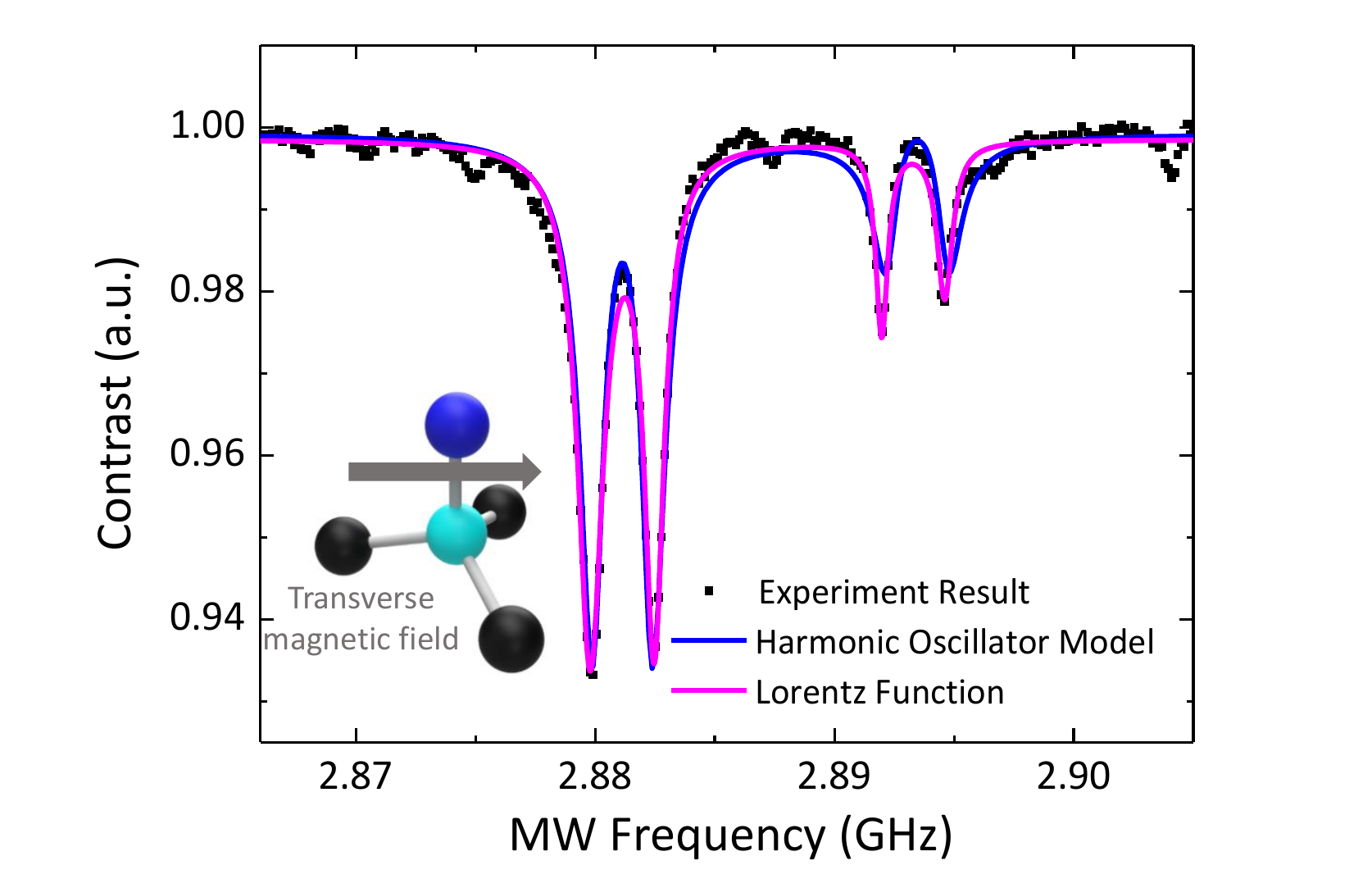}
  \caption{Experimental results of the CW-ODMR with the microwave and the radio-frequency fields under the effect of the transverse magnetic field. We sweep the microwave frequency from 2.866 GHz to 2.905 GHz. Also, we apply the RF fields of frequency in the MHz range. We fit experimental results with a Lorentzian function (pink) and with Eq. \eqref{E1} (blue). The orientation of the NV center and the directions of the magnetic fields (inset)}
  \label{RFdressedstateresult}
\end{figure}

We performed CW-ODMR with microwave and RF fields under the effect of a transverse magnetic field. CW-ODMR is performed by applying RF fields at ~MHz. Fig. \ref{RFdressedstateresult} shows results of the CW-ODMR experiment. Based on the four peaks observed in the CW-ODMR, we confirmed that RF-dressed states were generated. Although we tried to fit these results using a Lorentz function, this did not well reproduce experimental results because of the sharp dip structure in the spectrum. It is known that a combination of magnetic field noise and strain variations induces a sharp dip structure in CW-ODMR \cite{Saijo2018,Matsuzaki2016,Yamaguchi2019,Zhu2014,hayashi2018optimization}. Instead we use Eq. \eqref{E2} to fit the results and found a good agreement between experiment and theory in this case. In addition, by changing the laser power, the microwave intensity and the RF intensity, we perform the CW-ODMR. 

\begin{figure}[htbp]
  \centering
  \hspace*{-0.25cm}
  \includegraphics[width=9cm]{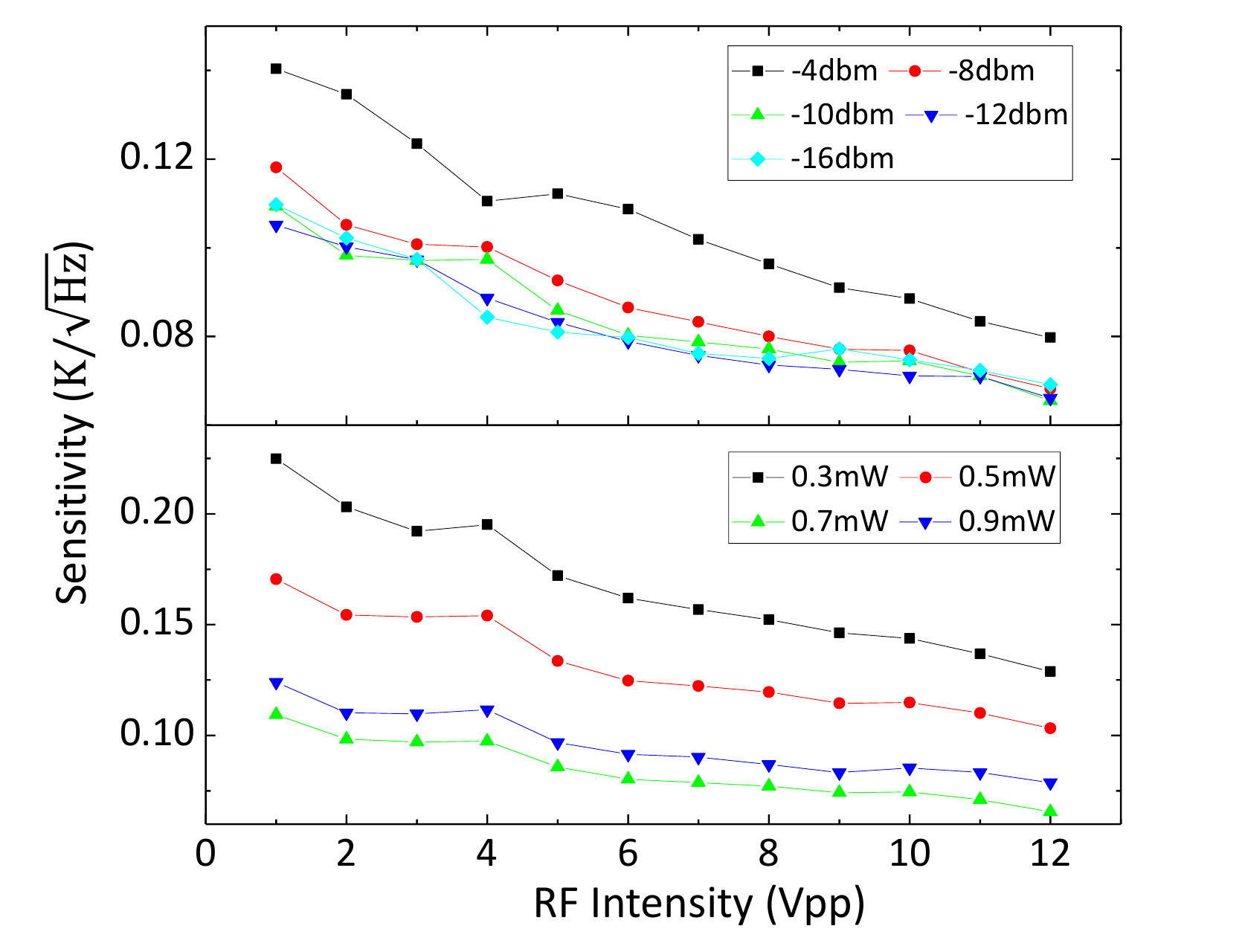}
  \caption{The sensitivity of the temperature sensing with the CW-ODMR. (a) Dependence of temperature sensitivity on the RF and microwave intensities. (b)Dependence of temperature sensitivity on the RF intensity and laser power.}
  \label{RFdressedstatesensitivity}
\end{figure}

From the fitting of CW-ODMR experimental results using Eq. \eqref{E2}, we can calculate the sensitivity as a temperature sensor. In Fig. \ref{RFdressedstatesensitivity}(a), we show how the sensitivity depends on the RF and MW intensity while we fix the laser power around 0.7 mW. However, as illustrated in Fig. \ref{RFdressedstatesensitivity}(b), we show how the laser power and the microwave intensity affect the sensitivity while we fix the microwave intensity around -10 dB. From these results, we determined the optimal conditions for the laser power, microwave intensity, and RF intensity, and obtained a sensitivity of 65.5 $\mathrm{m}\mathrm{K}/\sqrt{\mathrm{Hz}}$. 

\begin{figure}
  \centering
  \hspace*{-0.25cm}
  \includegraphics[width=9cm]{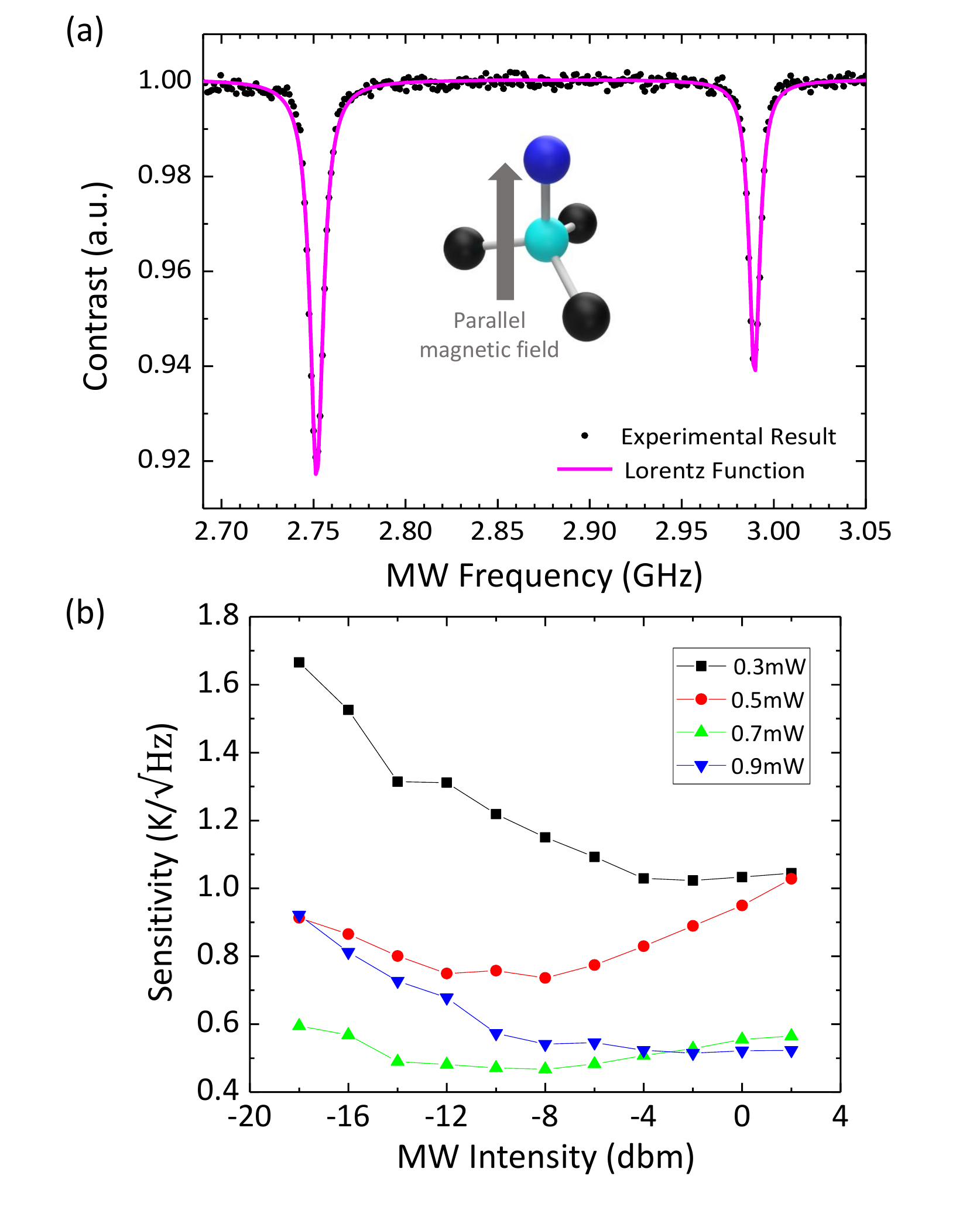}
  \caption{(a)The experimental result of CR-ODMR under the parallel magnetic fields. Microwave frequency is swept from 2.690 GHz to 3.050 GHz. We fit the experimental results with Lorentzian functions (pink). The orientation of the NV center and the directions of the magnetic fields (inset) (b)Dependency of temperature sensitivity on the microwave intensity and the laser power}
  \label{parallel}
\end{figure}

For comparison, we performed CW-ODMR with parallel magnetic fields. This is a typical approach for measuring temperature with CW-ODMR. Fig. \ref{parallel}(a) shows the experimental results of CW-ODMR. 
In addition, by changing the laser power and microwave intensity, we investigated the dependence of CW-ODMR on the laser power and microwave intensity. From the fitting of experimental results using a Lorentzian function, we can calculate the sensitivity of the temperature sensor. Fig. \ref{parallel}(b) shows the dependency of the sensitivity on the laser power and the MW intensity. From these results, we find an optimal condition for the laser power and MW intensity, and we obtained a sensitivity of approximately 467 $\mathrm{m}\mathrm{K}/\sqrt{\mathrm{Hz}}$ experimentally, which is seven times lower than that of our scheme.

\section{Discussion}
\begin{figure}[hbtp]
  \centering
  \hspace*{-0.25cm}
  \includegraphics[width=9cm]{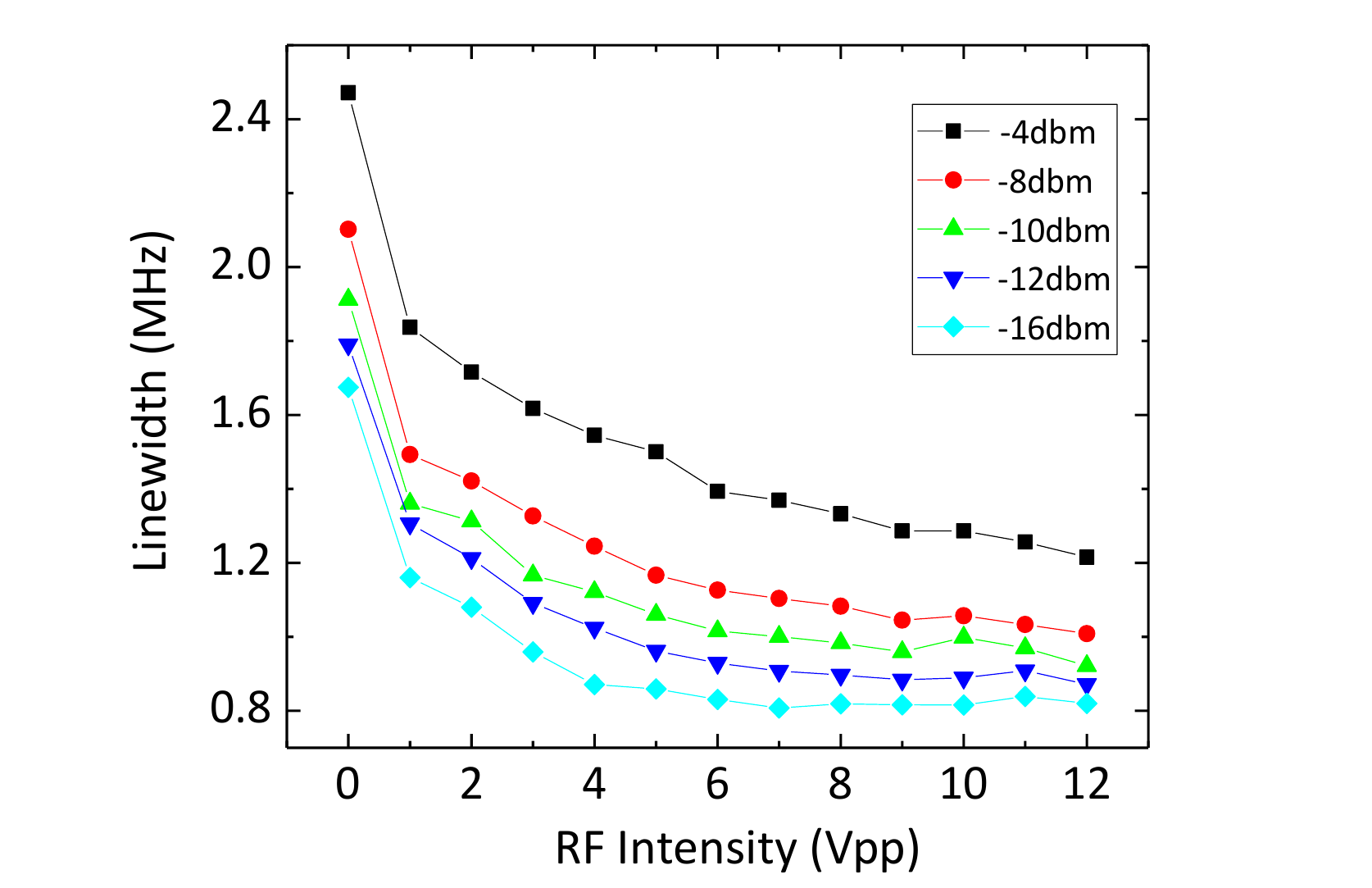}
  \caption{The linewidth of CR-ODMR with the microwave and the radio-frequency fields under the effect of the transverse magnetic field.}
  \label{width}
\end{figure}

Here we shall discuss the origin of sensitivity improvement observed with our scheme.
First, when we estimated the widths of peaks in CW-ODMR
for parallel (transverse) magnetic fields we had 
10.7 (2.47) MHz, 8.62 (2.10) MHz, 7.92 (1.91) MHz, 7.29 (1.79) MHz and 6.69 (1.67) MHz for the microwave intensity of -4dbm, -8dbm, -10dbm, -12dbm, -16dbm, respectively. 
This indicates that orthogonal magnetic fields suppress the noise, which contributes to the improvement in sensitivity.
Second, the RF-dressed states become robust to noise. As explained above, the effect of strain variations can be suppressed by applying an RF field.
As we increased the intensity of RF fields, the linewidth of the peak became narrower, as shown in Fig. \ref{width}. 

\section{Conclusion}
In this study, we described a novel method for measuring temperature using CW-ODMR with a quantum state dressed by RF fields under a transverse magnetic field. Because the RF fields suppress the effect of inhomogeneous broadening due strain variations, the sensitivity can be improved compared to the conventional scheme. Moreover, a sharp dip structure is observed in the CW-ODMR in our experiment, which also contributes to improving the sensitivity. We obtained an optimal temperature sensitivity of approximately 65.5 $\mathrm{m}\mathrm{K}/\sqrt{\mathrm{Hz}}$, which is seven times better than that of conventional schemes.

This work was supported by the Leading Initiative for Excellent Young Researchers MEXT Japan and JST presto (Grant No. JPMJPR1919), and Kanazawa University SAKIGAKE Project 2020. YM acknowledges the support of Grants-in-Aid for Scientific Research from the Ministry of Education, Culture, Sports, Science and Technology, Japan (20H05661).

\nocite{*}
%

\end{document}